BRIEF ARTICLE

# Automatic Quantification of Volumes and Biventricular Function in Cardiac Resonance. Validation of a New Artificial Intelligence Approach

*Cuantificación automática de los volúmenes y función de ambos ventrículos en resonancia cardíaca. Propuesta y evaluación de un método de inteligencia artificial*

ARIEL H. CURIALE[1,2,], MATÍAS E. CALANDRELLI[1,], LUCCA DELLAZOPPA[2,], MARIANO TREVISAN[3,], JORGE LUIS BOCIÁN[3,], JUAN PABLO BONIFACIO[3], GERMÁN MATO[1,2,4,]

**ABSTRACT**

**Background:** Artificial intelligence techniques have shown great potential in cardiology, especially in quantifying cardiac biventricular function, volume, mass, and ejection fraction (EF). However, its use in clinical practice is not straightforward due to its poor reproducibility with cases from daily practice, among other reasons.
**Objectives:** To validate a new artificial intelligence tool in order to quantify the cardiac biventricular function (volume, mass, and EF). To analyze its robustness in the clinical area, and the computational times compared with conventional methods.
**Methods:** A total of 189 patients were analyzed: 89 from a regional center and 100 from a public center. The method proposes two convolutional networks that include anatomical information of the heart to reduce classification errors.
**Results:** A high concordance (Pearson coefficient) was observed between manual quantification and the proposed quantification of cardiac function (0.98, 0.92, 0.96 and 0.8 for volumes and biventricular EF) in about 5 seconds per study.
**Conclusions:** This method quantifies biventricular function and volumes in seconds with an accuracy equivalent to that of a specialist.

**RESUMEN**

**Introducción:** Las técnicas de inteligencia artificial han demostrado tener un gran potencial en el área de la cardiología, especialmente para cuantificar la función cardíaca de ambos ventrículos, volumen, masa y fracción de eyección (FE). Sin embargo, su aplicación en la clínica no es directa, entre otros motivos por la poca reproducibilidad frente a casos de la práctica diaria.
**Objetivos:** Propuesta y evaluación de una nueva herramienta de inteligencia artificial para cuantificar la función cardíaca de ambos ventrículos (volumen, masa y FE). Estudiar su robustez para su uso en la clínica y analizar los tiempos de cómputo respecto a los métodos convencionales.
**Materiales y métodos:** Se analizaron en total 189 pacientes, 89 de un centro regional y 100 de un centro público. El método propuesto utiliza dos redes convolucionales incorporando información anatómica del corazón para reducir los errores de clasificación.
**Resultados:** Se observa una alta concordancia (coeficiente de Pearson) entre la cuantificación manual y la propuesta para cuantificar la función cardíaca (0,98, 0,92, 0,96 y 0,8 para los volúmenes y para la FE de ambos ventrículos) en tiempos cercanos a los 5 seg. por estudio.
**Conclusiones:** El método propuesto permite cuantificar los volúmenes y función de ambos ventrículos en segundos con una precisión comparable a la de un especialista.

**Palabras clave:** Aprendizaje profundo – Cardiopatías / Diagnóstico por imagen – Aplicación Open Source - Imagen por resonancia magnética

## INTRODUCTION

Data science applications have demonstrated great potential in the area of cardiology. This is the result of machine learning techniques —especially deep neural networks— that have successfully solved various classification, quantification and tissue detection tasks with an accuracy, in many cases, equivalent to interobserver error in the area of radiology. (1-3) A review recently published in the Journal of the American College of Cardiology (JACC) listed the promising achievements of artificial intelligence (AI) in cardiology, particularly in cardiovascular imaging. (4) However, the robustness and reproducibility of these techniques present a drawback (5) that makes it impossible to implement them in clinical practice. Therefore, there is a need to gain an objective understanding of the advantages and weaknesses when applying these techniques in the clinical setting. In this regard, we will analyze the robustness of deep neural networks to quantify cardiac function in terms of volumetric measurement and systolic function for both left and right ventricles in cardiac magnetic resonance imaging (CMRI).

A manual or semi-automatic quantification of the main cardiac structures (atria, ventricles, and myocardial tissue) normally used to diagnose and quantify various conditions such as infarction or hypertrophy, requires an intrinsically subjective, repetitive and laborious task that must be performed by a specialist. (6) This analysis takes 3-10 minutes per study, depending on the specialist's expertise. (7) Today, there are several applications for semi-automatic quantification, (8-10) but in clinical practice, detecting chambers and myocardial tissue is deficient. Most often, it involves manual intervention by a specialist, which is time-consuming. Therefore, the development of accurate, robust and subjectivity-free techniques continues to be an active area of research. In this respect, AI techniques seem to be the right tools to overcome these limitations and reduce quantification times to seconds.

This paper introduces a new AI application based on deep neural networks to automatically quantify cardiac function from the estimation of left and right ventricles and myocardial tissue in CMRI. Firstly, we have analyzed its robustness with multicenter data, with special emphasis on the effects and benefits observed when adjusting the application to CMRI of a medical center; and secondly, we have compared the time required by this method against the traditional manual method.

### METHODS

A retrospective, observational study was conducted in 89 patients (51±17 years, 67% men) with several conditions (left bundle branch block, acute myocardial infarction, dilated cardiomyopathy, hypertrophic cardiomyopathy, hypertensive cardiomyopathy), and normal cardiac function, from a health center in Bariloche (SSC). Images were obtained over a 2-year period with a Philips Intera 1.5T scanner. Chamber segmentation was performed by a specialist using the Segment software, which was also used to estimate the ejection fraction (EF), end-diastolic volume (EDV) and end-systolic volume (ESV) of both ventricles, and the LV myocardial mass.

A publicly available dataset, the Automatic Cardiac Diagnosis Challenge (ACDC) was used to assess the robustness of the proposed technique. (11) The database was made of 100 patients with conditions similar to those described above. Images were acquired over a 6-year period using two Siemens 1.5 and 3T scanners.

### Proposed method

CardIAc, the proposed application, was developed as an extension of the 3D Slicer software. (12) Figure 1 shows an example of the user interface and quantification for a patient from the SSC. The extension was developed to be used on a common computer, available today in any medical center.

Automatic quantification is performed with two convolutional neural networks (CNN) based on a U-Net architecture. (13) The first CNN is used to identify a region of interest (RoI) that includes the entire heart. This region is defined in 90 mm x 90 mm (∼128 x 128 pixels). Once the RoI around the heart is identified, a second neural network estimates the different structures of interest. The second neural network adds information of the anatomical structure of the heart following a variational autoencoder approach. (14) This way, errors in detecting anatomically non-plausible structures are reduced. Finally, biventricular volumes and EF and LV mass are quantified. Furthermore, if the user enters the patient's height and weight, volumes and mass are normalized using the body mass index.

### Statistical analysis

Data were collected as mean ± standard deviation, and differences were evaluated using a Student's t-test or t-test. In addition, the linear correlation between the proposed method and manual quantification was analyzed, and Pearson's correlation coefficient was calculated. Systematic errors and degree of concordance were assessed with a Bland and Altman analysis.

### Ethical considerations

The study was conducted in compliance with the National Data Protection Act No. 25 326, protecting patients' identity and personal data. All sensitive information was anonymized. The study was conducted in accordance with national and regional ethical standards, and its protocol was approved by the Ethics Committee in the province of Río Negro, Argentina.

### RESULTS

Three sets of experiments were carried out in order to analyze the accuracy and robustness of the proposed method. The aim of the first experiment was to study the accuracy of the proposed techniques using only public ACDC data, both for training and validation (Table 1).

The second experiment analyzed the accuracy of the proposed tool on daily practice data; in this regard, only SSC data provided by the local center were used (Table 2). In both cases, the accuracy of the proposed method was similar to the interobserver error for LVEF reported in the literature (2.7 ± 6.6%), (15) being 0.60 ± 4.77 % and 0.89 ± 4.55 % for ACDC and SSC respectively.

The third experiment studied the accuracy obtained when only a public database is used to train the proposed models (Table 3). It is important to note that the proposed tool was suitable to quantify the variables analyzed (Tables 1 and 2). However, when analyzing the robustness of the techniques on the data from the regional center, the accuracy decreases substantially compared to the results obtained when the proposed technique was trained on the regional SSC data. In particular, the error obtained when esti-

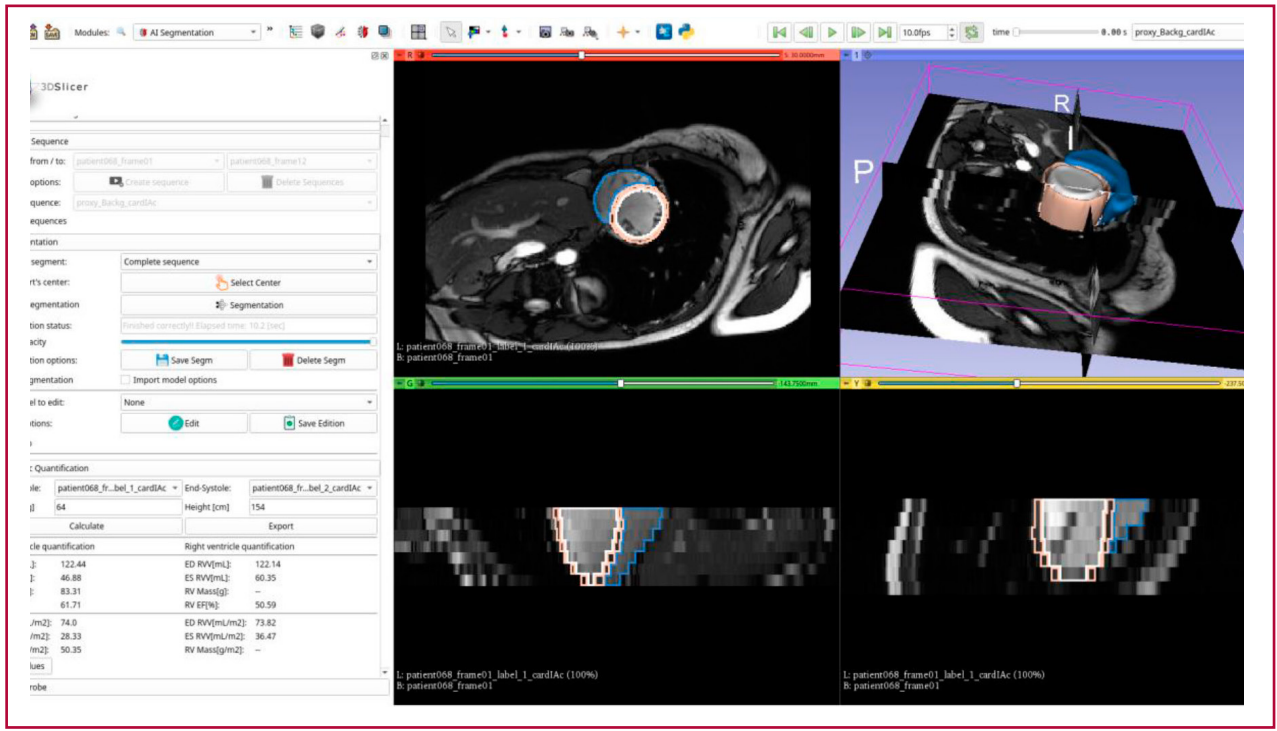

**Fig. 1**. Example of the proposed AI-based application within the 3D Slicer software

**Table 1**. Quantification of study variables for ACDC data (training and validation)

|  | Manual | AI | p | r | Bland-Altman |
|---|---|---|---|---|---|
| LV EDV [mml] | 164.61 ± 73.90 | 162.22 ± 70.93 | 0.01 | 0.99 | -2.39 (-20.08 a 15.29) |
| LV ESV [mml] | 99.06 ± 78.36 | 98.09 ± 75.74 | 0.314 | 0.99 | -0.97 (-19.07 a 17.76) |
| RV EDV [mml] | 152.99 ± 54.88 | 151.07 ± 54.15 | 0.142 | 0.97 | -1.92 (-27.20 a 23.36) |
| RV ESV [mml] | 85.88 ± 52.20 | 88.81 ± 55.52 | 0.091 | 0.95 | 2.92 (-30.50 a 36.35) |
| LVEF [%] | 46.38 ± 19.82 | 45.79 ± 19.69 | 0.217 | 0.97 | -0.60 (-9.94 a 8.75) |
| RVEF [%] | 46.73 ± 17.80 | 44.69 ± 17.89 | 0.018 | 0.89 | -2.04 (-18.60 a 14.52) |
| LV Mass [g] | 129.96 ± 50.75 | 129.36 ± 49.96 | 0.441 | 0.99 | -0.60 (-15.85 a 14.64) |

LV: Left ventricle. RV: Right ventricle. EDV: End-diastolic volume. ESV: End-systolic volume. EF: Ejection fraction. AI: Artificial inteligence.

**Table 2**. Quantification of study variables for SSC data (training and validation)

|  | Manual | AI | p | r | Bland-Altman |
|---|---|---|---|---|---|
| LV EDV [mml] | 166.97 ± 44.74 | 164.50 ± 41.87 | 0.05 | 0.98 | -1.65 (-15.56 a 12.26) |
| LV ESV [mml] | 74.32 ± 44.48 | 74.54 ± 43.70 | 0.83 | 0.98 | -0.13 (-11.73 a 11.98) |
| RV EDV [mml] | 149.80 ± 33.06 | 143.40 ± 35.03 | 0.002 | 0.92 | -6.41 (-43.17 a 19.61) |
| RV ESV [mml] | 61.89 ± 20.55 | 64.71 ± 22.45 | 0.003 | 0.95 | 2.82 (-13.96 a 19.61) |
| LVEF [%] | 58.02 ± 12.16 | 57.13 ± 15.31 | 0.07 | 0.96 | -0.89 (-9.82 a 8.04) |
| RVEF [%] | 59 ± 8.61 | 54.38 ± 14.45 | < 0.001 | 0.8 | -4.63 (-22.48 a 13.22) |
| LV Mass [g] | 141.11 ± 36.72 | 142.84 ± 35.08 | 0.21 | 0.97 | -1.14 (-12.32 a 14.59) |

LV: Left ventricle. RV: Right ventricle. EDV: End-diastolic volume. ESV: End-systolic volume. EF: Ejection fraction. AI: Artificial inteligence.

**Table 3.** Quantification of study variables for SSC data using ACDC (ACDC Training) and SSC (SSC Training) data as training

|  | Manual | ACDC Training | | SSC Training | |
| --- | --- | --- | --- | --- | --- |
|  |  | AI | r | AI | r |
| LV EDV [mml] | 166.97 ± 44.74 | 150.46 ± 38.52 | 0.98 | 164.50 ± 41.87 | 0.98 |
| LV ESV [mml] | 74.32 ± 44.48 | 73.35 ± 39.97 | 0.98 | 74.54 ± 43.70 | 0.98 |
| RV EDV [mml] | 149.80 ± 33.06 | 129.95 ± 35.98 | 0.92 | 143.40 ± 35.03 | 0.92 |
| RV ESV [mml] | 61.89 ± 20.55 | 69.35 ± 32.81 | 0.95 | 64.71 ± 22.45 | 0.95 |
| LVEF [%] | 58.02 ± 12.16 | 51.75 ± 20.94 | 0.96 | 57.13 ± 15.31 | 0.96 |
| RVEF [%] | 59 ± 8.61 | 46.76 ± 19.22 | 0.8 | 54.38 ± 14.45 | 0.8 |
| LV Mass [g] | 141.11 ± 36.72 | 106.10 ± 30.17 | 0.97 | 142.84 ± 35.08 | 0.97 |

LV: Left ventricle. RV: Right ventricle. EDV: End-diastolic volume. ESV: End-systolic volume. EF: Ejection fraction. AI: Artificial inteligence.

mating LVEF increases to 6.27 ± 12.13 % (more than double), much higher than the interobserver error presented above, and even unacceptable for clinical practice. The proposed method's processing time was 5 seconds per patient and 452.7 sec ± 191.9 for the manual operator, with a difference between times of 447.7 sec (95% CI 407.6 - 487.8, p < 0.001).

### DISCUSSION
This study demonstrated that AI techniques —in particular deep neural networks— have a promising future in Cardiology, especially for classifying various heart structures in CMRI, significantly reducing computational times (5 seconds vs. 7-8 min per study). Furthermore, and this is one of its main contributions, the study objectively showed the dependence of these techniques on training data, representing a major limitation particularly in the clinical setting. Supervised machine learning, such as those proposed in this paper, are strongly dependent on the amount and generality of the data used for training. In this regard, the differences showed in Table 3 are mainly due to the following factors: (1) the quality of the images used in public competitions, particularly in ACDC, is often higher than those in clinical practice; (2) it is common to rule out studies that include artifacts in CMR images; and (3) it is common to rule out areas where the myocardium is incomplete, for example at the base or directly in the apex. As a result of these factors, a significant difference is observed when using ACDC data during training, and then to assess and quantify SSC data, which does not exclude any part of the myocardium.

### Limitations
The proposed tool needs refinement for proper determination of RV volumes, as the RV structure is more complex than that of the LV. Further multicenter studies are required to increase and validate the accuracy of the application, particularly in centers with different equipments and patients with different heart conditions.

### CONCLUSIONS
To the best of our knowledge, ours is the first study in Argentina to introduce a user-friendly tool to quantify cardiac function using AI techniques, which in turn, can be adapted to the needs of different national centers to obtain accurate quantification in a few seconds and in real-world cases.

### Conflicts of interest
None declared.
(See authors' conflict of interests forms on the web/Additional material.)